# Decomposition of waves in time series of data related to Covid-19

**Amelia Carolina Sparavigna**
Department of Applied Science and Technology, Politecnico di Torino

*Here it is proposed a decomposition in components of the "waves" which appear in the time series of data related to Covid-19 pandemic. The decomposition is based on functions of κ-statistics; in particular the κ-Weibull is used. Fitted data are those of the "waves" ranging from August 2020 to April 2021 in the United Kingdom, from September 2020 to May 2021 in Ireland, and from September 2020 to June 2021 in Italy. For the United Kingdom, the time series of daily infection shows a wave composed by two peaks. Among the many factors involved in the spread of infection, it seems that, in driving the onset of the second peak, the main role was played by the emergence of Alpha variant of Sars-Cov-2. Therefore, the proposed decomposition of waves in the time series is applied to study the role of Alpha variant in the spread of infection. For what concerns data from Ireland, if we decompose the analysed "wave" in three components, we can observed that for one of them the role of the Alpha variant was relevant for sure. In the case of Italy, the effect is less evident, because it was spread over a larger time interval.*

**Introduction**
It seems that no formal definition exists for the "wave" of a pandemic, however the term is associated to a rising number of Covid-19 cases, which is characterized by a specific peak and then by a decline. In Ref.1 it is told that "Public health scientists first began using this term [wave] to describe different peaks and valleys of infections during influenza outbreaks in the late 1800s and the 1918-1929 Spanish flu." The Reference is also stressing that each "wave" has a different feature and can impact different populations, even within the same country. From data such as those given by www.worldometers.info, we can see that the trend of Covid-19 infection is generally given, from the second wave, by the composition of two or more peaks. For instance, in Italy we had a first wave which was characterized by an isolated peak, but from September 2020, the time series of data related to pandemic were characterized by a composition of some peaks (more than two peaks). In this framework, we can try to analyse data to distinguish the peaks in each "wave" - or the wavelets in a wave train - in order to have an instrument able of determining the onset of a specific component. In this manner, having the onset-time it is possible to identify what caused the surge of further infections. Let us note that, sometimes, the term "deconvolution" is applied to a process of decomposing peaks that overlap with each other.

The aim of the discussion here proposed is that of applying the κ-statistics to evaluate the onset of the peaks in the "waves". In particular we will compare the results obtained from fitting the data given by www.worldometers.info and coronalevel.com with information coming from Our World in Data, www.ourworldindata.org, concerning the variants of the virus. In the following, the function used for analysis is that proposed in [2], that is the κ-Weibull probability distribution function. The data from time-series will be analysed using a 7-day moving average. The decomposition of waves here proposed is mainly applied to study the role of Alpha variant in the spread of infection.

**Method - Weibull and κ-Weibull pdf**
In the proposed method we will use the κ-Weibull function. In [2], the κ-Weibull probability distribution function (pdf) is described by:

$$f_\kappa(t) = \frac{\alpha \beta t^{\alpha-1}}{\sqrt{1+\kappa^2 \beta^2 t^{2\alpha}}} \exp_\kappa(-\beta t^\alpha) \quad (1)$$

where the κ-exponential is defined in the following manner:

$$\exp_\kappa(u) = \left(\sqrt{1+\kappa^2 u^2} + \kappa u\right)^{1/\kappa} \quad (2)$$

Eq. 1 is describing the κ-Weibull function. Parameters α,β are related to the shape and scale indexes of Weibull distribution, whereas κ is the index of κ-distribution, that is the distribution introduced by G. Kaniadakis, Politecnico di Torino, in [3,4].
In the formalism of [5], the Weibull probability density function (pdf) is defined as:

$$f(t|B,C,D) = \frac{B}{C}\left(\frac{t-D}{C}\right)^{(B-1)} e^{-\left(\frac{t-D}{C}\right)^B} \quad (3)$$

where $B>0$, $C>0$, $-\infty<D<\infty$, $t>D$.
In Weibull pdf, symbol $t$ is representing the random variable. Here, that is in the analysis of time-series, it is the elapsed time. Parameter D is the threshold, which is therefore representing the minimum value of time. B is the shape parameter, which controls the overall shape of the probability density function. Its value usually ranges between 0.5 and 8.0 [5].
The Weibull distribution includes other useful distributions [5]. If $B=1$, we have the exponential distribution. For $B=2$, we have the Rayleigh distribution. For $B=2.5$ and $B=3.6$, the Weibull distribution approximates the lognormal distribution and the normal distribution respectively. The scale parameter C changes the scale of the probability density function along the time axis (that is from days to months or from hours to days). It does not change the actual shape of the distribution [5]. Parameter C is known as the characteristic life. In [5], it is stressed that "No matter what the shape, 63.2% of the population fails by t = C+D". It is also told that "Some authors use 1/C instead of C as the scale parameter".
In [6], we can find discussed and defined the κ-Weibull. In the formalism of the given reference:

$$f_\kappa = \frac{m}{x_s}\left(\frac{x}{x_s}\right)^{m-1} \frac{\exp_\kappa\left(-[x/x_s]^m\right)}{\sqrt{1+\kappa^2(x/x_2)^{2m}}} \quad (4)$$

In (4), $x$ is the random variable. In the formalism of [5], with time and threshold:

$$f_\kappa(t|B,C,D) = \frac{B}{C}\left(\frac{t-D}{C}\right)^{B-1} \frac{\exp_\kappa\{-[(t-D)/C]^B\}}{\sqrt{1+\kappa^2((t-D)/C)^{2B}}} \quad (5)$$

Let us put $\alpha = B$, $\gamma = 1/C$, $T = D$. (5) becomes:

$$f_\kappa(t|\alpha,\gamma,T) = \alpha\gamma\gamma^{\alpha-1}(t-T)^{\alpha-1}\frac{\exp_\kappa\{-\gamma^\alpha(t-T)^\alpha\}}{\sqrt{1+\kappa^2\gamma^{2\alpha}(t-T)^{2\alpha}}} \quad (6)$$

Then, using $\beta = \gamma^\alpha$:

$$f_k(t|\alpha,\beta,T) = \frac{\alpha\beta(t-T)^{\alpha-1}}{\sqrt{1+\kappa^2\beta^2(t-T)^{2\alpha}}}\exp_\kappa(-\beta(t-T)^\alpha) \quad (7)$$

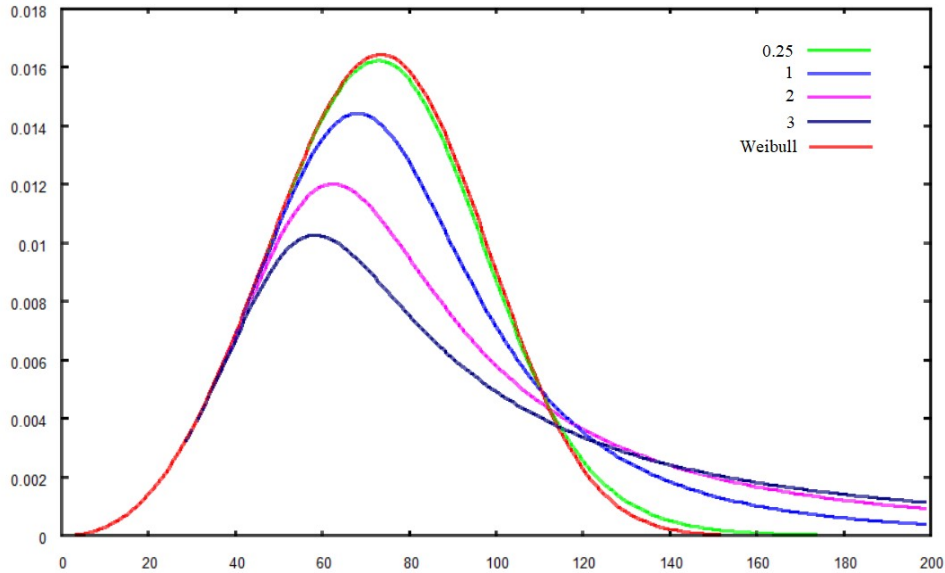

Figure 1 – Comparing Weibull and κ-Weibull. The Weibull pdf is given in red. Parameters are $\alpha = 3.5$, $\beta = 2.0\times 10^{-7}$, and T=0. The κ-Weibull curves have different κ values: 0.25, 1, 2 and 3.

Figure 1 shows the comparison of Weibull pdf with that of κ-Weibull. We can see that the value of κ parameter is strongly affecting the tail of the distribution. Increasing the value the tail becomes a long tail, that is, a portion of the distribution having many occurrences far from the head of the distribution. For the choice of the parameters used in the Figure 1, let us consider the Figure 2, which is involving data from China, already considered in [2]. Data are a courtesy by coronalevel.com. This web site is providing data from the Johns Hopkins University, Center for Systems Science and Engineering (CSSE), which can be used to have the specific geographic detail.
In the following Figure 2, the fit is made by means of κ-Weibull pdf.

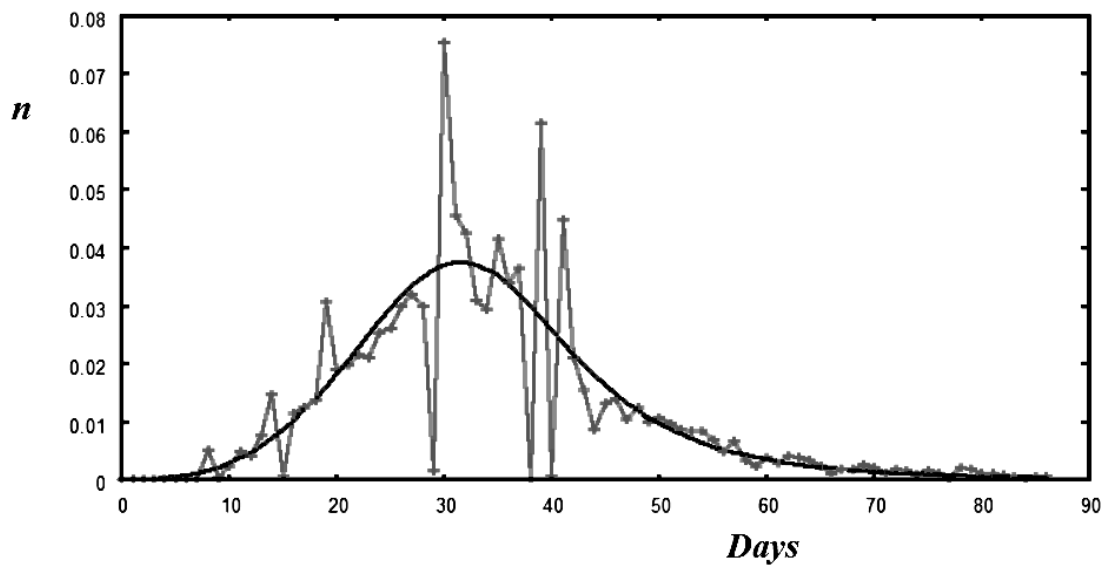

Figure 2 – Number of daily cases (victims), China, divided by the total number of cases, observed during the first wave. Parameters for the fit (red curve) are $\kappa = 0.8$, $\alpha = 4.0$, $\beta = 5.5 \times 10^{-7}$. Residual sum of squares $6.59 \times 10^{-3}$.

A further comparison of the role of κ parameter in the cases that we will discuss here, can be made with respect to Gaussian pdfs. Let us consider again the data concerning the first wave of pandemic in China. From Figure 2, it could appear Gaussian functions also suitable to fit data, however a difference exists and it is regarding the tail.
For fitting data by means of a Gaussian pdf, let us use program *Fityk* https://fityk.nieto.pl , developed by M. Wojdyr [7]. Comparing Figures 2 and 3, we can observe that κ-Weibull is properly describing the tail, whereas Gaussian pdf underestimates it.

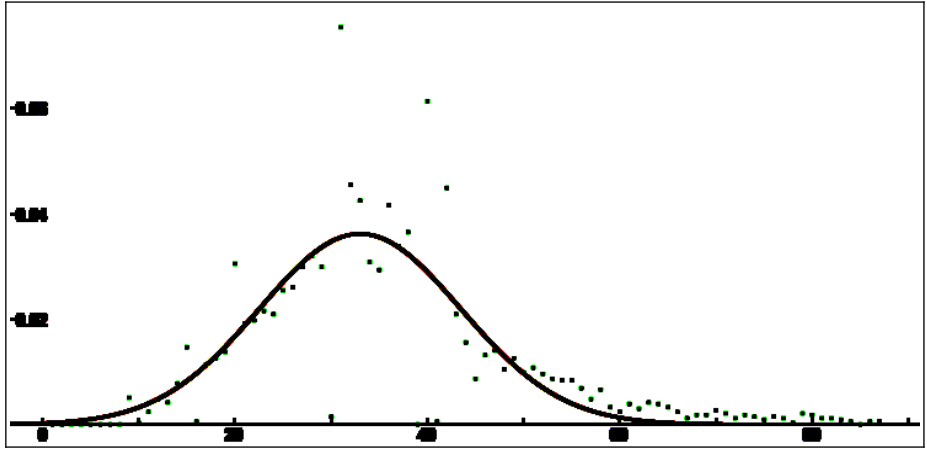

Figure 3 – Fit obtained by means of program *Fityk* https://fityk.nieto.pl , with a Gaussian pdf. Image Courtesy software Fityk. The weighted sum of squared residuals (WSSR) is 0.0067.

**Method – Mixture density**

In [2], where the first wave of Covid-19 had been analysed for China, Italy, Germany, Spain and United Kingdom, we have seen that (1) is properly fitting the data of time-series. Now, let us consider the case of a time-series which is characterized by the composition of two peaks. In particular we will analyse the data (courtesy www.worldometers.info and coronalevel.com) related to the surge of the second wave in the United Kingdom, which will be shown in the Figures 4 and 5. In the figures, the daily number of cases (7-day moving average) is divided by the total number of cases, observed in the considered time period (in days). If we consider the case of these data, let us try to use two functions (7), to fit the time-series, in the following form:

$$f = f_1 + f_2 = \xi f_{\kappa_1}(t|\alpha_1,\beta_1,T_1) + (1-\xi) f_{\kappa_2}(t|\alpha_2,\beta_2,T_2) \quad (8)$$

Parameter $\xi$, the mixing parameter, ranging from zero to 1, is used to generalize the addition of peaks, as proposed for the Weibull distribution [8]. It is also a rough manner to consider the fact that the set of population, involved by pandemic, changed for sure during the considered time period (we will further discuss this point).

In the case that we have three peaks, then (8) becomes:

$$f = f_1 + f_2 + f_3 = \xi_1 f_{\kappa_1}(t|\alpha_1,\beta_1,T_1) + \xi_2 f_{\kappa_2}(t|\alpha_2,\beta_2,T_2) + \xi_3 f_{\kappa_3}(t|\alpha_3,\beta_3,T_3) \quad (9)$$

In (9), we must have $\xi_1 + \xi_2 + \xi_3 = 1$ .

Being a finite sum, the mixture is known as a finite mixture, and the density is the "mixture density". Usually, "mixture densities" can be used to model a statistical population with subpopulations. Each component is related to a subpopulations, and its weight is proportional to the given subpopulation in the overall population. In the following discussion related to data concerning the United Kingdom, we have two subpopulations: we could guess one population infected by the Sars-CoV-2 virus (earlier strains), and the other subpopulation by its Alpha variant.

The mixture densities (8) and (9) have been used in [9], for the study of individual incomes.

**The United Kingdom**

As previously told, in [2] the first wave of Covid-19 had been analysed for China, Italy, Germany, Spain and the United Kingdom, using (1) which is properly fitting the data of time- series. Here we consider data (courtesy www.worldometers.info , 7-day moving average) from the surge of the second wave in the United Kingdom. The time-series is characterized by the composition of two peaks.

In the analysis, the daily number of cases is divided by the total number of cases, observed in the considered time period (in days). The κ-Weibull pdf is used in the form (7) in a mixture (8):

$$f_k(t|\alpha,\beta,T) = \frac{\alpha\beta(t-T)^{\alpha-1}}{\sqrt{1+\kappa^2\beta^2(t-T)^{2\alpha}}} \exp_\kappa(-\beta(t-T)^\alpha)$$

$$f = f_1 + f_2 = \xi f_{\kappa_1}(t|\alpha_1,\beta_1,T_1) + (1-\xi)f_{\kappa_2}(t|\alpha_2,\beta_2,T_2)$$

Fitted data given in the Figure 4 are those ranging from August 2020 to April 2021 in the United Kingdom. Numbers are related to the daily new infections caused by the virus Sars-CoV-2. The data baseline has been shifted of 1180 cases.

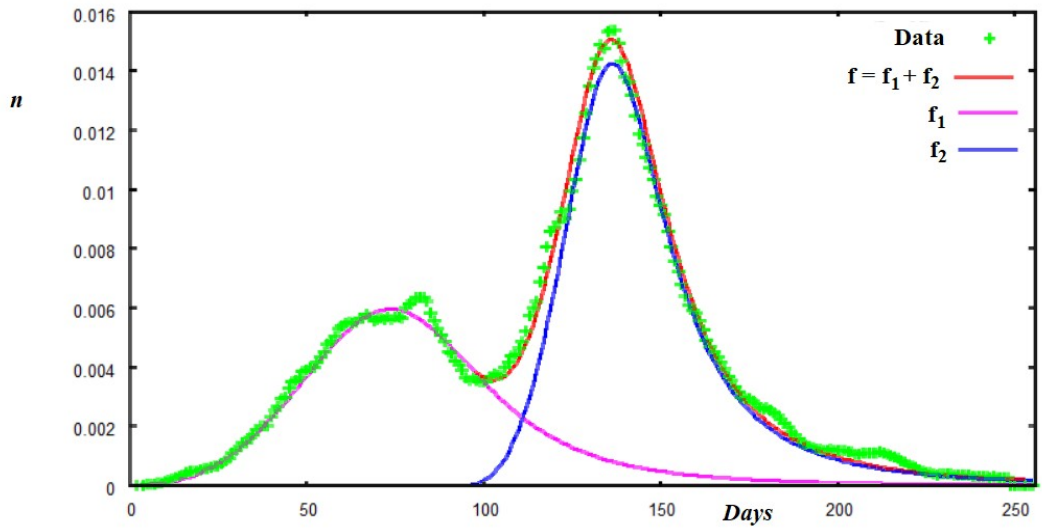

Figure 4 – Number of daily cases (infections) divided by the total number of cases, observed in the considered time period. Data from the United Kingdom. Parameters for the fit (red curve) are $\kappa_1=0.72$, $\kappa_2=1.25$ , $\alpha_1=3.62$, $\alpha_2=3.85$ , $\beta_1=1.0\times10^{-7}$, $\beta_2=2.6\times10^{-7}$ , $T_1=0$, $T_2=94$ . $\xi=0.4$ . The first day was August 20, 2020. The day 94 was November 21, 2020.

Using the κ-Weibull we find that the threshold time for the second peak - time $T_2$ - was November 21, 2020. This is the onset of the second peak in the Figure 4. A concomitant spread of the Alpha variant of Sars-CoV-2 can be observed in data given by Our World in Data. Among the many factors involved in the spread of infection, the Alpha variant of the virus seems the one which had the main role.

As previously told, the decomposition is giving the onset of the second peak on November 21, 2020. From the web site Our World in Data, using this [LINK](LINK) we can evidence that, on 23 November 2020 for instance, the percentage of Alpha variant of Sars-CoV-2 in the United Kingdom was of 7.8 %. Using the data reported in the Table 1, we can argue that this variant had a leading role in the onset of the second largest peak of infections. Actually, the large increase in percentage of Alpha variant is the main feature of the second component of the wave train.

TABLE 1

1.96 % – November 9, 2020
----- $T_2$ -----
7.8 % – November 23, 2020
12.51 % – December 7, 2020
40.24 % – December 21, 2020
63.60 % – January 4, 2021
75.81 % – January 11, 2021
83.56 % – January 25, 2021
92.04 % – February 8, 2021
95.97 % – February 22, 2021

Table 1 – Percentage of Alpha Variant in analysed sequences for United Kingdom according to Our World in Data. Many, many thanks to this site and people involved in it.

Let us note that the web site, Our World in Data, tells that data are "The share of analysed sequences in the last two weeks that correspond to each variant group. This share may not reflect the complete breakdown of cases since only a fraction of all cases are sequenced".

Fitted data given in the Figure 5 are, as those given in the Figure 4, ranging from August 2020 to April 2021 in the United Kingdom. Numbers are related to victims of the virus. Using the κ-Weibull we find the threshold time for the second peak at day 94, which is corresponding to the first week of December 2020. The increase of infections, due to spread of the new variant, had the awful consequence of increasing victims of Covid-19.

arXiv:2111.03636

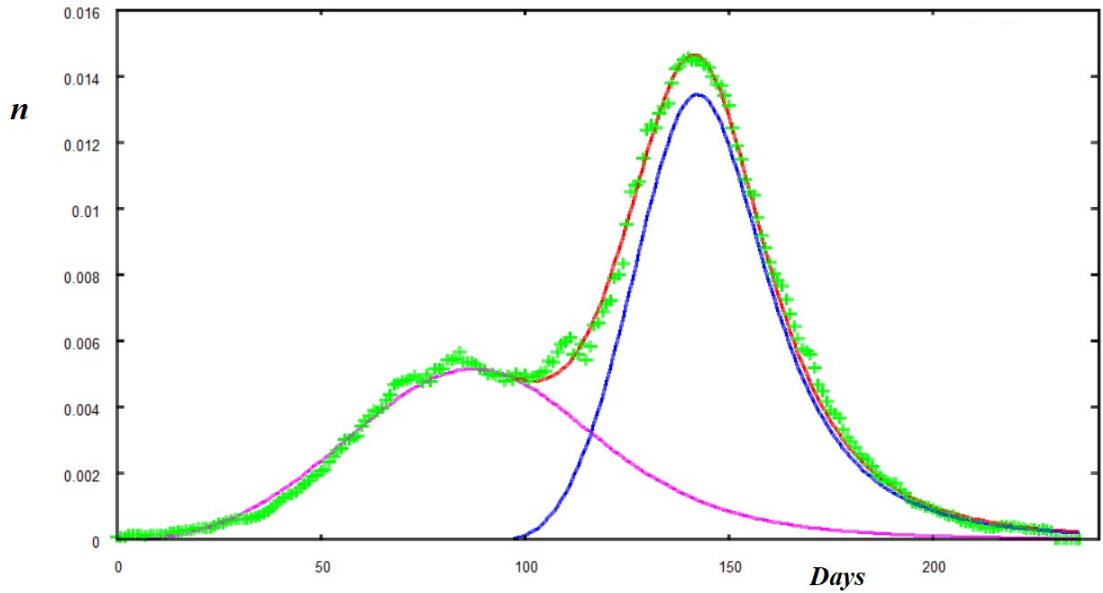

Figure 5 – Number of daily cases (victims) divided by the total number of cases. Parameters for the fit (red curve) are $\kappa_1=0.52$, $\kappa_2=0.90$, $\alpha_1=3.50$, $\alpha_2=3.75$, $\beta_1=1.0\times 10^{-7}$, $\beta_2=2.6\times 10^{-7}$, $T_1=0$, $T_2=94$. $\xi=0.4$.

**Alpha variant**

The Alpha variant is also known as lineage B.1.1.7 (other names are given by [wikidata.org](wikidata.org)). It is one of the several variants of concern (VOC), recognized by the World Health Organization. In [10], it is told that B.1.1.7 had been first detected in Kent on 20 September 2020. The variant spread quickly across the United Kingdom and its presence was reported before the start of the second English lockdown (5 November 2020). "By the end of that lockdown (2 December 2020), B.1.1.7 was widespread throughout the UK" [10]. In [11], we can find also told by Verity Hill, co-author of [10], that "The Alpha variant began by spreading mostly within London and the South East, even during the November lockdown in England. Once this was lifted, it spread rapidly across the country, as human movement increased significantly".

In [10], the authors were "able to trace the origins of the Alpha back to a point source in the South East of England" [11], and the county of Kent is mentioned. Therefore, in the framework of this analysis of the daily new cases, it can be interesting to observe the trend for Kent too. The web site coronalevel.com is providing data from Kent. In fact, we can see, thanks to this site, the behaviour of the daily number of new infections for each county of England. Here in the following Figure 6, the waves of infections in Kent from May 2020.

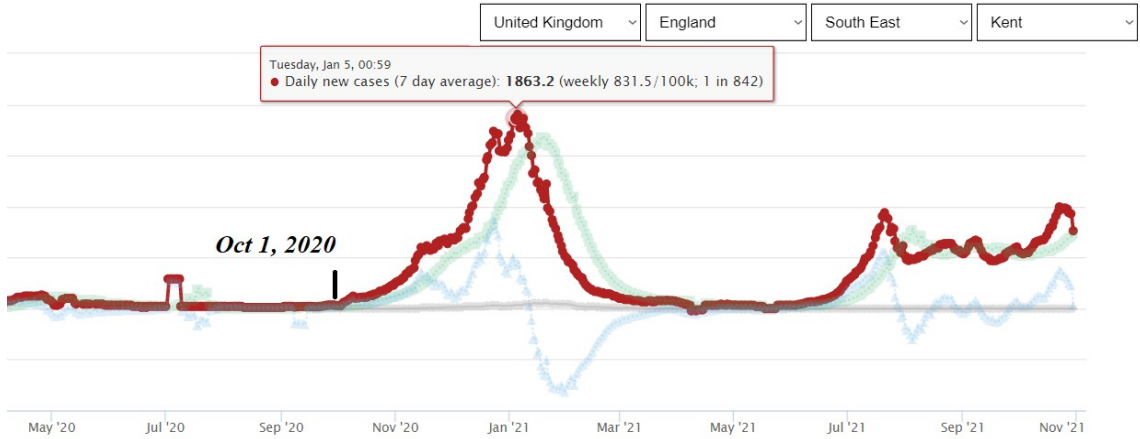

Figure 6 – Daily new cases in Kent – Data and graph are a courtesy by coronalevel.com - https://coronalevel.com/United_Kingdom/England/South_East/Kent/

The wave seems starting from October 1, 2020.
We can also compare the wave in Kent, with that observed in London (Figure 7). With the aim of distinguishing the role of Alpha variant in the case of London, we can apply the same analysis as made for Fig.4. The result is proposed in the Figure 8.

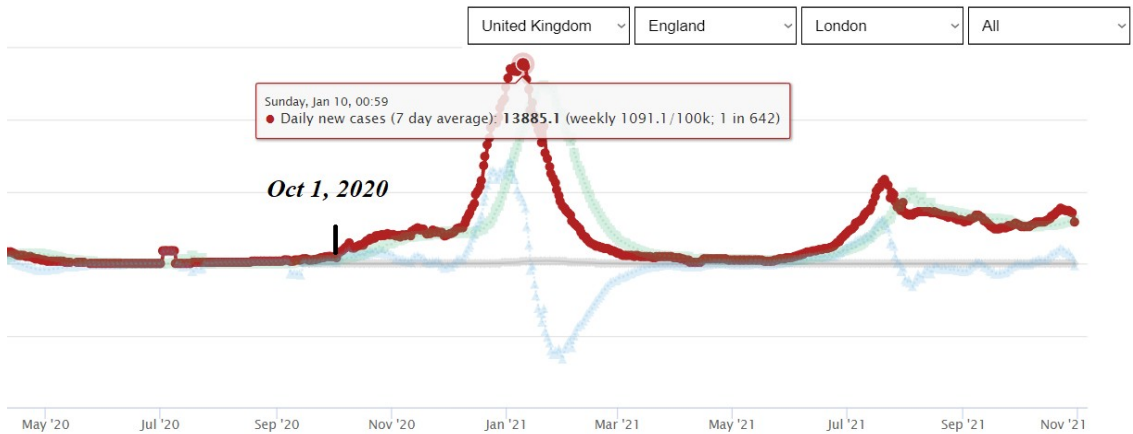

Figure 7 – Daily new cases in London – Data and graph are a courtesy by coronalevel.com - https://coronalevel.com/United_Kingdom/England/London/

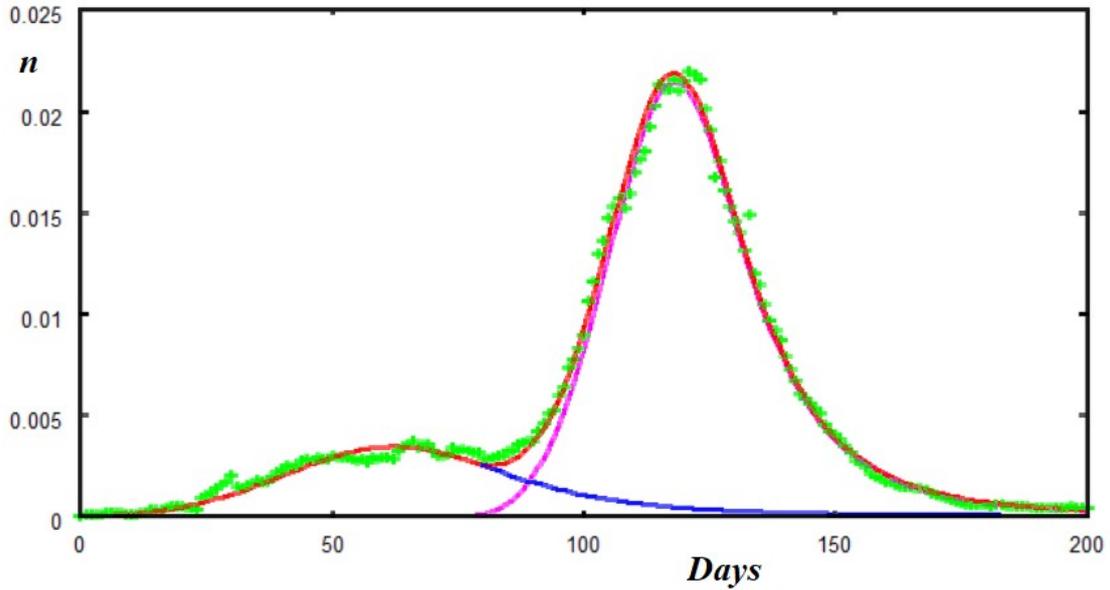

Figure 8 – Number of daily cases (infections), London, divided by the total number of cases, observed in the considered time period. Parameters for the fit (red curve) are
$\kappa_1=0.7$, $\kappa_2=0.9$ , $\alpha_1=3.5$, $\alpha_2=3.9$ , $\beta_1=3.0\times10^{-7}$, $\beta_2=2.6\times10^{-7}$ ,
$T_1=0$, $T_2=77$ . $\xi=0.2$ . Residual sum of squares $5.47\times10^{-5}$ .
The first day was September 10, 2020. The day 77 was November 25, 2020.

For the threshold time, November 25, we have an agreement with the threshold previously found of November 21 from the data of the United Kingdom. Comparing Figure 7 with Figure 9, we can see that the behaviour of the wave in London is almost the same as that of data from the region of South East of England.

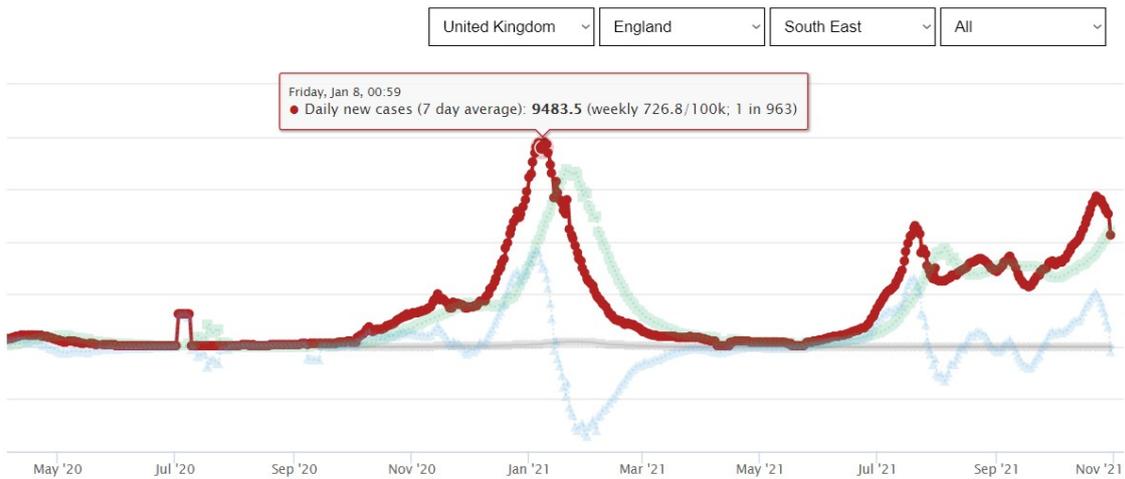

Figure 9 – Daily new cases in the region of South-East of England – Data and graph are a courtesy by coronalevel.com -  https://coronalevel.com/United_Kingdom/England/South_East/

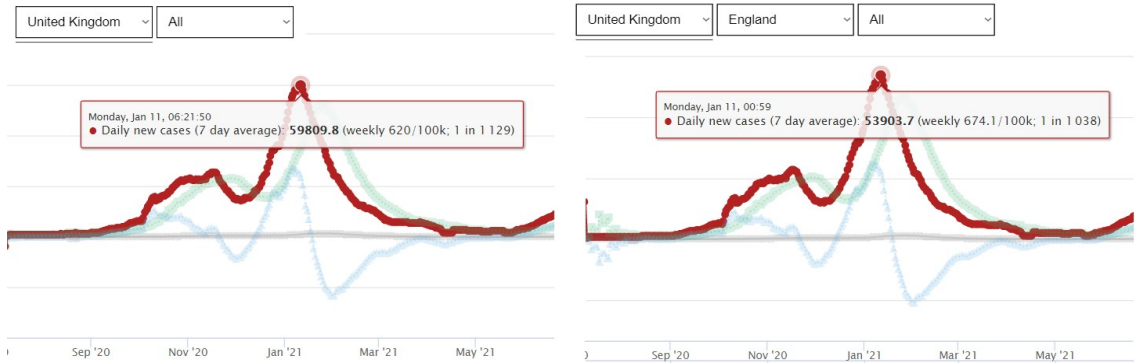

Figure 10 – Daily new cases in the United Kingdom (left) and in England (right)– Data and graphs are a courtesy by coronalevel.com

We can repeat the comparison of the behaviours of the second "wave" for England and the United Kingdom. From the Figure 10, we can see that the behaviour is almost the same.

**During a lockdown**
The data in the previous Figures show what we can find told in [11], that the Alpha variant began its spread "even during the November lockdown in England". In wikipedia.org, it is told that England entered the second national lockdown on 5 November 2020. "On 23 November, the government published a new enhanced tier system which applied in England following the end of the second lockdown period on 2 December. On 16 December, [the Prime Minister Boris] Johnson said that restrictions would be relaxed for five days over the Christmas period. That same day, the Health Secretary Matt Hancock announced that a new COVID-19 strain had been discovered, which was named VUI-202012/01 [that is the Alpha variant]. On 20 December Johnson said that the planned Christmas relaxations had been cancelled for London and South East England and limited to a single day for the rest of England as a result of the discovery of the strain". England entered the third lockdown from 5 January 2021".
It is clear that lockdowns or restrictions are able to stop the spread of infections. However, the second component of the "wave" appeared during a lockdown, when the first component of the wave was decreasing. It seems therefore that the specific lockdown, which started on November 5, 2020, was able to contrast the previous strain of the virus, not the Alpha variant. For this reason, we can argue that the second component of the "wave" was mainly driven by the new variant, which produces a more infectious disease.

**Some recent literature**
Several researches on the Alpha variant were recently published.
In [12], we can find an article published on April 9, 2021. It is told that the "rapid spread of VOC 202012/01 [Alpha variant] suggests that *it transmits more efficiently from person to person than preexisting variants* of SARS-CoV-2. This could lead to global surges in COVID-19 hospitalizations and deaths, so there is an urgent need to estimate how much more quickly VOC 202012/01 spreads, whether it is associated with greater or lesser severity of disease, and what control measures might be effective in mitigating its impact". The authors used "social contact and mobility data, as well as demographic indicators linked to SARS-CoV-2 community testing data in England, to assess whether the spread of the new variant may be an artifact of higher baseline transmission rates in certain geographical areas or among specific demographic subpopulations". The authors used complementary statistical analyses and mathematical models to estimate the transmissibility of the new variant. They extended a mathematical model "that has been extensively used to forecast COVID-19 dynamics in the UK to consider two competing SARS-CoV-2 lineages: VOC 202012/01 and preexisting variants". By fitting their model to a variety of data sources, the authors assessed different hypotheses concerning the spread of the new variant.

In Ref. [13], it is told that in "England, the emergence and rapid spread of the SARS-CoV-2 Alpha (B.1.1.7 or Kent) variant in November 2020 led to a second wave of cases, hospitalizations and deaths, resulting in prolonged national lockdown including school closures". *Ref. [13] studied infection among students and staff of school*. Authors tell that "Early in the pandemic, schools in England were closed as part of national lockdown in March 2020 and only partially reopened for some school years in June 2020. Since September 2020, however, all schools fully reopened for in-person teaching. As per national guidance, face masks and face coverings were not recommended in classrooms, but staff and children in secondary schools were advised to wear them in communal areas outside the classroom if physical distancing was difficult to maintain. Cases in adults and children increased throughout September and October 2020 and a second national lockdown was imposed for adults from 05 November to 02 December 2020, whilst keeping all schools open. Cases fell rapidly first in adults and then in children even though all schools remained fully open at the time." However, we have seen before that "By the end of that lockdown (2 December 2020), B.1.1.7 was widespread throughout the UK" [6]. In any case, let us note that, in the time series proposed above, the peak was reached when England entered the third lockdown from 5 January 2021.

In Ref.[14], the alpha variant is told being "associated with *higher transmissibility than wild type* virus". Sars-CoV-2 B.1.1.7 became the dominant variant of the virus in England by January 2021. The authors of [14], "aimed to describe the severity of the alpha variant in terms of the pathway of disease from testing positive to hospital admission and death". After the descriptions of methods and results, the authors concluded that "The SARS-CoV-2 alpha variant is associated with an *increased risk* of both hospitalisation and mortality than wild-type virus".

Here just three references are shortly discussed. But a search on November 3, 2021, by means of Google Scholar https://scholar.google.com/scholar?as_ylo=2020&q=+SARS-CoV-2+Alpha+B.1.1.7 provides 2660 results. Let us stress that among the results, we can find also articles where the Alpha variant is not the main subject of the research but it is just mentioned for comparison.

After the analysis of the role of Alpha variant in the spread of infection due to Sars-

CoV-2 in the United Kingdom, we can move to other countries to evidence the presence of different time thresholds.

**Ireland**

Let us consider the data of daily new infections in Ireland from September 1, 2020, to May 1, 2021. The data from time-series are analysed using a 7-day moving average. Using two peaks as in the previous analysis for the United Kingdom, we can obtain a fit as in the Figure 11. It is evident that at least one peak more is required for the analysis of the largest component of the wave.

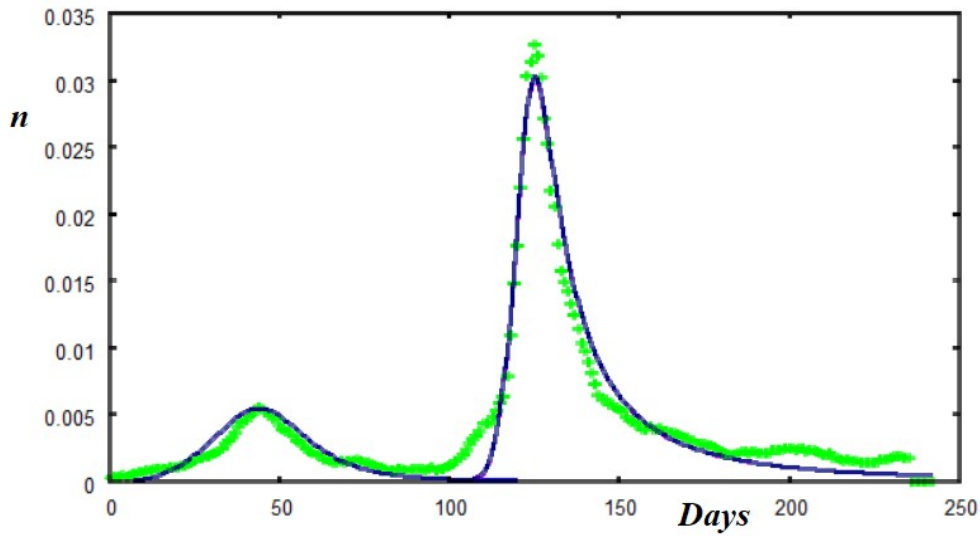

Figure 11 – Number of daily cases of infections (Ireland) divided by the total number of cases. Parameters are $\kappa_1=0.8$, $\kappa_2=3.25$, $\alpha_1=4.0$, $\alpha_2=4.7$, $\beta_1=1.5\times10^{-7}$, $\beta_2=2.0\times10^{-7}$, $T_1=0$, $T_2=105$. $\xi=0.2$.

Just for proposing the method, let us consider only three peaks. Then, we use for the fit of data mixture (8):

$$f=f_1+f_2+f_3=\xi_1 f_{\kappa_1}(t|\alpha_1,\beta_1,T_1)+\xi_2 f_{\kappa_2}(t|\alpha_2,\beta_2,T_2)+\xi_3 f_{\kappa_3}(t|\alpha_3,\beta_3,T_3)$$

Here, we must have $\xi_1+\xi_2+\xi_3=1$.

In the Figure 12, a fit for data from Ireland is proposed based on this model. The data baseline has been shifted of 90 cases (the same in the Figure 11).

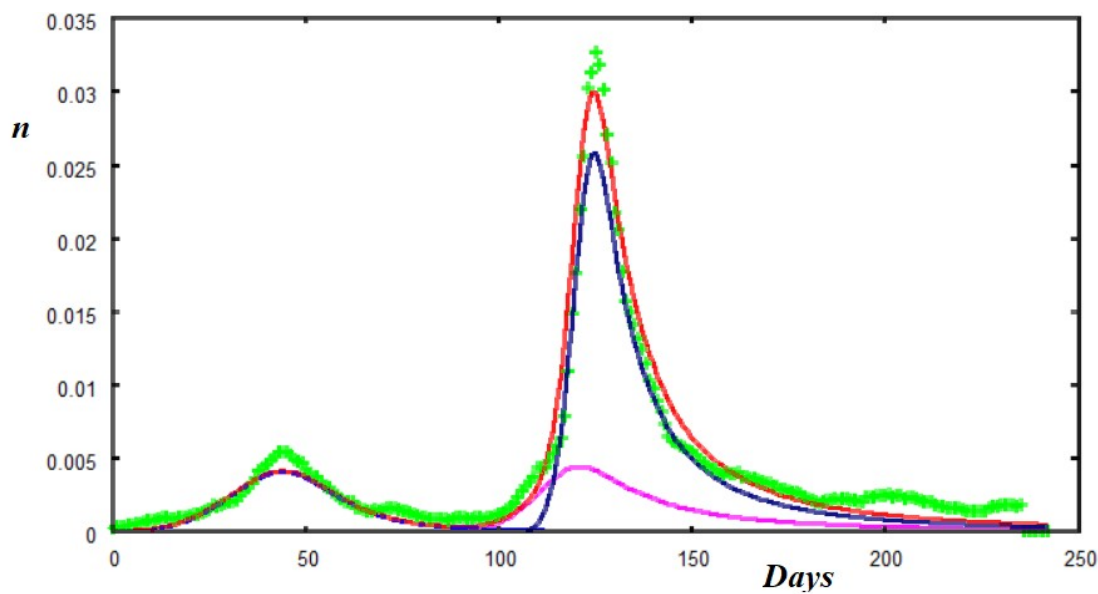

Figure 12 – Number of daily cases (infections), Ireland, divided by the total number of cases, observed in the considered time period. Parameters for the fit (red curve) are
$\kappa_1 = 0.8$, $\kappa_2 = 2.6$, $\kappa_3 = 3.2$ , $\alpha_1 = 4.00$, $\alpha_2 = 4.20$, $\alpha_3 = 4.75$ ,
$\beta_1 = 1.5 \times 10^{-7}$, $\beta_2 = 1.0 \times 10^{-7}$, $\beta_3 = 2.0 \times 10^{-7}$ , $T_1 = 0$, $T_2 = 85$, $T_3 = 105$ .
$\xi_1 = 0.15$, $\xi_2 = 0.20$, $\xi_3 = 0.65$ . The first day was September 1, 2020. The day 85 was November 24, 2020. The day 105 was December 14, 2020.

As from the Figure 11, we have two components with thresholds $T_2 = 85$, $T_3 = 105$ . Threshold $T_2 = 85$ corresponds to November 24, 2020, and threshold $T_3 = 105$ to December 14, 2020. From the web site Our World in Data, we can have the percentage of Alpha variant of Sars-CoV-2 . Data are reported in the Table 2; again we can argue that this variant had a leading role in the onset of the largest peak of infections in Ireland. Actually, it seems that the second peak was also influenced by the presence of the Alfa variant.

TABLE 2
0.00 % – October 26, 2020
2.86 % – November 9, 2020
0.00 % – November 23, 2020
----- $T_2$ -----
0.00 % – December 7, 2020
----- $T_3$ -----
11.00 % – December 21, 2020
46.15 % – January 4, 2021
52.14 % – January 11, 2021
61.39 % – January 25, 2021
81.76 % – February 8, 2021
86.64 % – February 22, 2021
93.68 % – March 8, 2021

As told before for the United Kingdom, it is clear that lockdowns or restrictions are able to stop the spread of infections. The restrictions applied to Ireland were able to control and reduce the spread of infections until November 24, 2020. But, at this time, the second component of the "wave" appeared, followed by the third on December 14. Again, it seems that the specific restrictions, which were able to contrast the previous strain of the virus, were not enough to control the Alpha variant. In this manner, we can argue that the largest peak was driven by the Alpha variant. This peak is made of two components, one with threshold on November 24, 2020, the other with threshold on December 14, 2020. Then, in this case, the role of the Alpha variant is more evident for the second component. The first could have been driven by this variant, but also by other causes, such as relaxation of some restrictions.

In [wikipedia.org](wikipedia.org), the largest peak in Figures 11 and 12 is defined as the "Third Wave: December 2020–July 2021". It is told that "On 17 December, the National Public Health Emergency Team recommended to the Government of Ireland that the period of relaxed COVID-19 restrictions from 18 December be shortened to the end of the year as COVID-19 cases rise. - On 21 December, speaking at a COVID-19 press briefing, the Chair of the NPHET Irish Epidemiological Modelling Advisory Group Philip Nolan announced that a third wave of COVID-19 in Ireland was clearly underway. - On 22 December, the Government of Ireland agreed to move the entire country to Level 5 lockdown restrictions with a number of adjustments from Christmas Eve until 12 January 2021 at the earliest. … On 23 December, in a statement from the National Public Health Emergency Team, the Chair of the NPHET Coronavirus Expert Advisory Group Cillian de Gascun announced that the new variant of COVID-19 in the United Kingdom was now present in the Republic of Ireland, based on a selection of samples analysed from the weekend. Two days later on 25 December (Christmas Day), Chief Medical Officer Tony Holohan officially confirmed that the new UK variant of COVID-19 had been detected in the Republic of Ireland by whole genome sequencing at the National Virus Reference Laboratory in University College Dublin. By week 2 of 2021, the variant had become the dominant strain in Ireland".

**Italy**
Let us consider data concerning Italy from September 2020 to June 2021. In this case we need at least three peaks. Just for the proposal of the method, let us use again the mixture of three functions (9):

$$f = f_1 + f_2 + f_3 = \xi_1 f_{\kappa_1}(t|\alpha_1, \beta_1, T_1) + \xi_2 f_{\kappa_2}(t|\alpha_2, \beta_2, T_2) + \xi_3 f_{\kappa_3}(t|\alpha_3, \beta_3, T_3)$$

$$\xi_1 + \xi_2 + \xi_3 = 1$$

In the Figure 13, a fitting is given. The data baseline has been shifted of 1237 cases. However, it is clear that to have a better result, more than three functions are required, such as a further adjustment of the baseline.

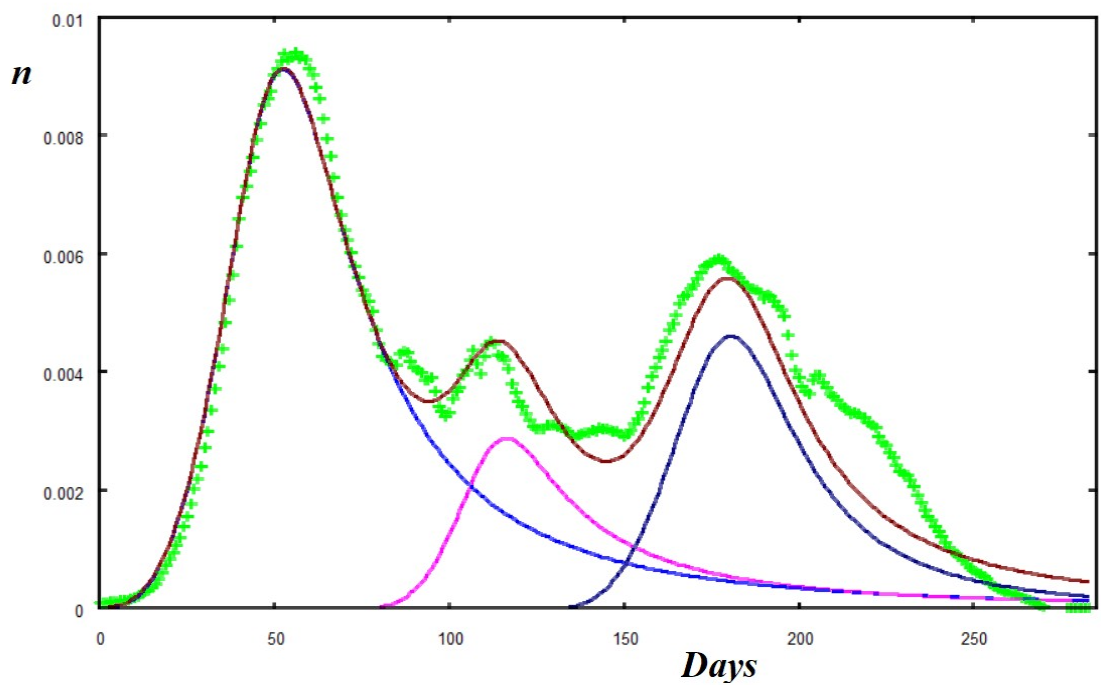

Figure 13 – Number of daily cases (infections), Italy, divided by the total number of cases, observed in the considered time period. Parameters for fit (red curve) are
$\kappa_1=2.1$, $\kappa_2=3.2$, $\kappa_3=1.5$, $\alpha_1=4.00$, $\alpha_2=3.98$, $\alpha_3=3.80$,
$\beta_1=0.55\times10^{-7}$, $\beta_2=1.10\times10^{-7}$, $\beta_3=1.50\times10^{-7}$, $T_1=0$, $T_2=75$, $T_3=130$.
$\xi_1=0.57$, $\xi_2=0.18$, $\xi_3=0.25$. The first day was September 15, 2020. The day 75 was November 29, 2020. The day 130 was January 22, 2021.

Our World in Data tells that the percentage of Alpha variant in Italy was:

TABLE 3
0.77 % – October 12, 2020
1.78 % – October 26, 2020
0.51 % – November 9, 2020
1.16 % – November 23, 2020
----- $T_2$ -----
0.94 % – December 7, 2020
4.82 % – December 21, 2020
20.63 % – January 4, 2021
25.75 % – January 11, 2021
----- $T_3$ -----
25.34 % – January 25, 2021
59.81 % – February 8, 2021
44.74 % February 22, 2021
69.38 % – March 8, 2021

If we use a decomposition of data concerning the pandemic wave from September 2020 to June 2021 in three components, comparing times $T_2$, $T_3$ with data giving the percentage of the Alpha variant in analysed sequences of Sars-CoV-2 as in Table 3, we can argue that this variant had a relevant role in the second "wave" in Italy.

For Ireland, if we decompose the analysed "wave" in three components, we can observed that for one of them the role of the Alpha variant was relevant for sure. In the case of Italy, the effect is less evident, because it was spread over a larger time interval.

Here the restrictions in Italy from October 2020 to March 2021, from it.wikipedia .
"Con una legge, dall'8 ottobre 2020 diventa obbligatorio l'uso della mascherina sia nei luoghi all'aperto sia al chiuso. … il parlamento italiano approva una legge in vigore dal 13 ottobre che limita le possibilità di assembramento con regole precise per attività quali ristoranti, cinema, teatri, competizioni sportive e feste. Il 7 novembre viene imposto il coprifuoco, generalmente tra le 22:00 e le 5:00, per cui è vietato ogni spostamento. Il 26 ottobre 2020 vengono nuovamente chiusi centri sportivi, cinema, teatri, musei, sale giochi e sale scommesse, e la frequentazione di bar e ristoranti è consentita fino alle 22:00. Con il DPCM del 3 novembre 2020, … [e poi rettificato], le Regioni italiane vengono raggruppate in tre tipi di scenari epidemiologici diversi. Viene istituito in tutta la nazione un coprifuoco dalle 22.00 alle 5.00, si ordina la chiusura dei centri commerciali nel fine settimana e il ricorso alla didattica a distanza per le scuole superiori. … Con un decreto-legge del 2 dicembre si impongono inoltre restrizioni agli spostamenti fra Regioni nel periodo delle festività natalizie, in particolare a partire dal 21 dicembre 2020 e fino al 6 gennaio 2021; a queste restrizioni si aggiungono quelle del decreto-legge del 18 dicembre, che fra il 24 dicembre 2020 e il 6 gennaio 2021 prevede il passaggio dell'intero territorio nazionale in zona rossa nei giorni festivi e prefestivi, e in zona arancione nelle giornate feriali. Le stesse misure vengono prorogate con il decreto-legge del 5 gennaio 2021, che stavolta prevede una zona gialla nazionale nei giorni feriali e una zona arancione nei giorni prefestivi e festivi, fino al 15 gennaio. A partire dall'11 gennaio riprende la didattica in presenza nelle scuole superiori al 50-75% (tranne nelle zone rosse). Il divieto di spostamento fra Regioni viene prorogato con ulteriori decreti-legge fino al 25 aprile 2021. Il decreto n. 2 del 14 gennaio 2021 istituisce una "zona bianca" per le aree a basso rischio di contagio. Con il DPCM del 14 gennaio 2021 si dispone la riapertura dei musei nei giorni feriali in zona bianca e gialla e il divieto di asporto per i bar dopo le ore 18.00, mentre con il DPCM del 2 marzo 2021 si dispone la chiusura di scuole, parrucchieri ed estetisti nelle zone rosse, e in zona bianca e gialla la riapertura dei musei anche nel fine settimana, e di cinema e teatri a partire dal 27 marzo 2021."

**Discussion about the Mixture density**

The κ-Weibull function can be used in a finite sum to have a finite mixture and a related "mixture density". As previously told, a "mixture densities" can be used to model a statistical population with subpopulations. Each component is related to a subpopulations, and its weight is proportional to the given subpopulation in the overall population. In the discussion related to data regarding the United Kingdom and London, we have two subpopulations. We could guess one population infected by the Sars-CoV-2 virus (earlier strains), and the other subpopulation by its Alpha variant. In the case of Fig.13 (Italy), the mixture density was based on three subpopulations, where two are linked to the presence of Alpha variant. In this case, it would be better to find another feature concerning the spread of infection to distinguish the two Alpha variant subpopulations, that is the second and third distributions.

Again, as we have made for China and one peak, let us use a mixture of Gaussian function for comparison in the case of two peaks. Once more, for fitting data by means of a Gaussian pdf, let us use program *Fityk* https://fityk.nieto.pl . Comparing Figures 8 and 14 (data from London), we can observe that κ-Weibull is properly describing the tail, whereas Gaussian pdf underestimates it.

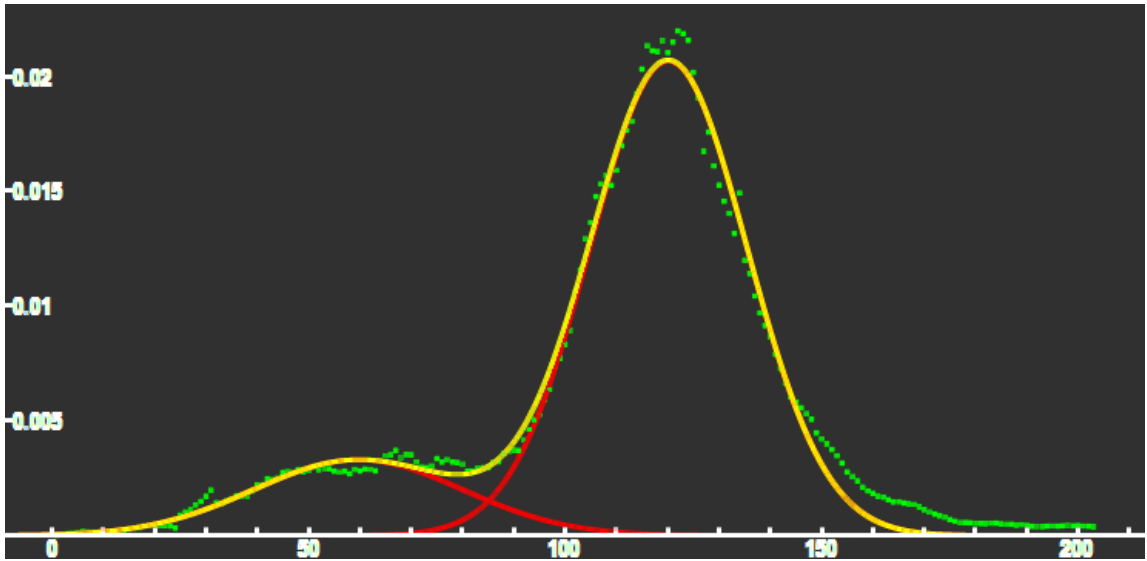

Fig. 14 - Two Gaussian functions - Image Courtesy software Fityk. WSSR: $9.\times 10^{-5}$ . Note that the long tail is not properly described.

As we have seen before in the case of data from China, the κ-Weibull pdf is able to provide a more precise description of the tail than the use of a Gaussian one.

Let us consider the case of a mixture of pdfs. In the Figure 8 we have proposed a fit with κ-Weibull of the daily new cases in London. The results was obtained using two functions, having $\alpha_1 = 3.5$, $\alpha_2 = 3.9$ . One of these parameters is close to the value of 3.6, for which the Weibull pdf approximates the normal (Gaussian) pdf. Therefore, a question about the results we can obtain fitting the data by means of two Gaussian pdfs is well posed. Using the program *Fityk* https://fityk.nieto.pl , the result is given in the Fig. 14 (the fit concerns data of the Figure 8 previously given). The weighted sum of squared residuals (WSSR), also called chi-square, is given in the caption too. Again, as stressed before, we have a difference for the long tail. Then, let us use three Gaussian functions in the same software to fit the tail. The result is given in the Fig. 15.

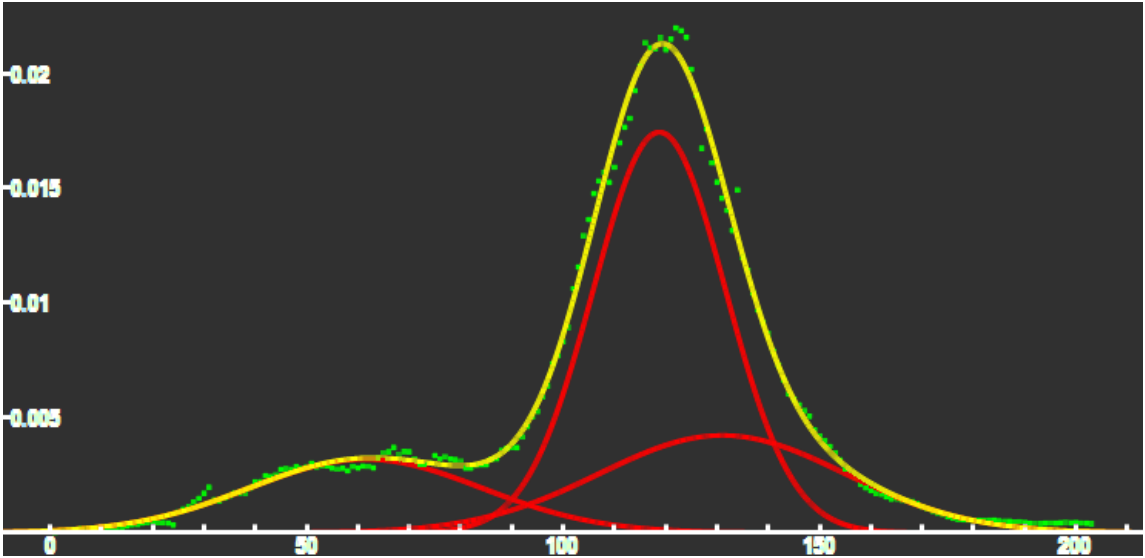

Fig. 15 - Three Gaussian functions - Image Courtesy software Fityk. WSSR: $3.\times 10^{-5}$ .

If we use three functions, we have not the possibility to define a unique threshold time for the second peak in the wave.

**A note on non-extensivity** - In [8], it is told that in contrast with the Weibull model, the hazard function of the κ-Weibull is non-extensive. The same is also well-known for the entropy in κ-statistics [15]. In fact, if we have two independent systems $A$ and $B$, entropy is given by [16]:

$$S_\kappa^{A\cup B} = S_\kappa^A \, I_\kappa^B + S_\kappa^B \, I_\kappa^A \quad (12)$$

The formula (12) is representing a generalized additivity of entropy. The entropy $S_\kappa$ and function $I_\kappa$ can be expressed by means of Euler's infinite product expansions [17], which can be useful in numerical calculations.
In (4) and (5), actually, we have a generalized additivity of pdf. Further refinements of this additivity, in the framework of κ-statistics, are under consideration.

—————

Here we have proposed the use of κ-Weibull to decompose the peaks in the time series linked to Covid-19 pandemic - The method is based on a mixture density – We have compared the onset-time of peaks with the percentage of Alpha variant in the analysed sequences - The use of more than three functions is required for the analysis of the pandemic in many cases.